\documentclass[runningheads]{llncs}
\usepackage[T1]{fontenc}
\usepackage{graphicx}
\usepackage{xcolor}
\usepackage{orcidlink}
\usepackage{svg}
\begin{document}
\title{Deep histological synthesis from mass spectrometry imaging for multimodal registration}
\titlerunning{Deep histological synthesis from mass spectrometry imaging}
\author{Kimberley M. Bird\inst{1}\orcidlink{0009-0003-2395-5803} \and
Xujiong Ye\inst{2}\orcidlink{0000-0003-0115-0724} \and
Alan M. Race\inst{3}\orcidlink{0000-0001-8996-2641} \and James M. Brown\inst{1}\orcidlink{0000-0001-7636-4554}}
\authorrunning{K.M. Bird et al.}
\institute{University of Lincoln, Lincoln, Lincoln, U.K.\\
\and University of Exeter, Exeter, U.K.\\
\and AstraZeneca Computational Pathology GmbH, Munich, Germany\\}

% \email{\{15589926@students.,jamesbrown@\}lincoln.ac.uk}\\
% \email{X.Ye2@exeter.ac.uk}\\
% \email{alan.race1@astrazeneca.com}}
%
\maketitle              % typeset the header of the contribution
\begin{abstract}
Registration of histological and mass spectrometry imaging (MSI) allows for more precise identification of structural changes and chemical interactions in tissue. With histology and MSI having entirely different image formation processes and dimensionalities, registration of the two modalities remains an ongoing challenge. This work proposes a solution that synthesises histological images from MSI, using a pix2pix model, to effectively enable unimodal registration. Preliminary results show promising synthetic histology images with limited artifacts, achieving increases in mutual information (MI) and structural similarity index measures (SSIM) of $+0.924$ and $+0.419$, respectively, compared to a baseline U-Net model. Our source code is available on GitHub: \url{https://github.com/kimberley/MIUA2025}.
% The abstract should briefly summarize the contents of the paper in
%\keywords{deep synthesis \and histology \and mass spectrometry imaging.}
\end{abstract}
\section{Introduction}
Mass spectrometry imaging (MSI) can visualise and localise thousands of known and unknown biomarkers by their molecular mass at a cellular level. Additionally, histological images, such as hematoxylin and eosin (H\&E) stained tissues, reveal valuable structural information regarding cell morphology and tissue types. Spatial registration of these modalities allows for precise localisation of biomolecules within structures of interest. However, due to the destructive nature of MSI and H\&E, images are typically adjacent tissue slices, increasing the risk of non-homologous information between two sections \cite{buchberger2017}. Furthermore, registering thousands of MSI channels to a regular 3-channel histology image is intractable, leading researchers to reduce MSI dimensionality using conventional techniques, such as principal component analysis (PCA) \cite{race2013}, t-distributed stochastic neighbor embedding (t-SNE) \cite{abdelmoula2016}, and uniform manifold approximation and projection (UMAP) \cite{race2021}.\\
Recent studies have proposed the use of generative models within digital pathology for a variety of purposes: converting H\&E to multiplex immunohistochemistry (IHC) \cite{bian2025}, producing synthetic IHC from masks for downstream segmentation \cite{winter2025}, synthetic multi-staining of microscopy tissue \cite{shi2023}, and generating IHC from H\&E \cite{liu2022}. This paper proposes the use of pix2pix \cite{isola2017} to generate synthetic histology from MSI to enable downstream unimodal registration. The pix2pix model was compared to a U-Net \cite{ronneberger2015} model trained on histology, and evaluated based on their perceived quality and similarity to real histology. To our knowledge, this is the first work to perform image-to-image translation in the generation of paired histological and MSI images.
\section{Materials and Methods}
We used a publicly available dataset of breast tissue desorption electrospray ionization MSI (DESI-MSI) and histology images \cite{guenther2015}. The MSI data was acquired at a spatial resolution of $100\mu m$, and the histology samples were stained using H\&E, totalling 111 pairs. The MSI data was pre-processed by applying interpolation rebinning \cite{race2016} to standardise m/z values across the entire dataset. After rebinning, all spectra were summed and the top 50 peaks were selected. Subsequently, image pairs were rescaled and resized (using black or white padding for histology where necessary) to match dimensions, followed by affine control point registration. For registration, a three-channel representation of the MSI was used (via either PCA, t-SNE, or UMAP) depending on its visual clarity. For the U-Net experiment, image pairs were split into $32\times32$ patches. The dataset was split into 70\% training, 10\% validation, and 20\% testing.\\
The U-Net model was developed using Python 3.8.10 and implemented in Keras 2.11.0, using an Intel Core i5-6500 CPU and NVIDIA RTX 2080Ti 11GB GPU. The model had minimal modifications from \cite{ronneberger2015} (batch size=64, learning rate=0.01, optimiser=Adam, activation=sigmoid, loss=mean squared error). The pix2pix model was implemented in PyTorch 2.4.1, changing several hyperparameters: input image size=$512\times512$, learning rate=0.00002, and L1 penalty=200. All models were trained until they converged. Performance was evaluated using mutual information (MI) and structural similarity index measure (SSIM).
% A visualisation of the full pipeline is shown in Fig. \ref{fig1}.
%
% Create a figure that shows a pipeline of (1) preprocessing, (2) control point registration, (3) synthesis, and (4) output.
% \begin{figure}
% \includegraphics[width=\textwidth]{fig1.eps}
% \caption{\color{red}The full pipeline illustrating ...\color{black}} \label{fig1}
% \end{figure}
%
\section{Results and Discussion}
\begin{figure}[h]
\begin{center}
\includegraphics[width=250px]{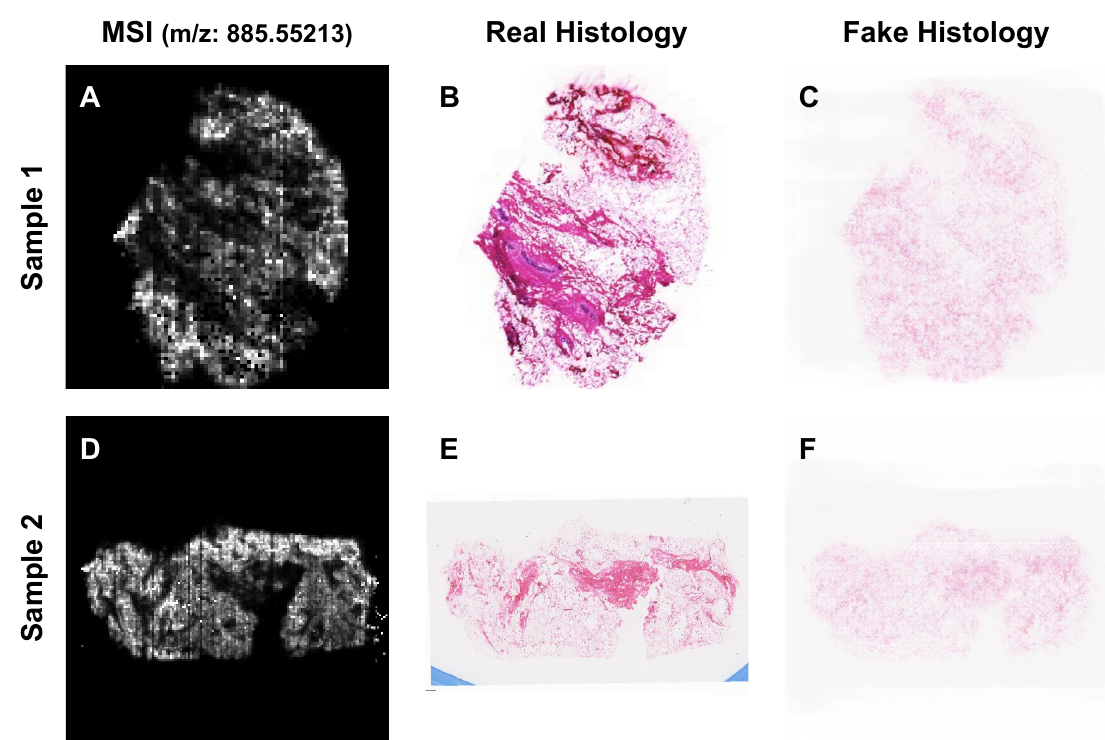}
\end{center}
\caption{Two examples of synthesised histology tissue samples (C+F) compared to their corresponding MSI (A+D) and real histology (B+E).} \label{fig1}
\end{figure}

Overall, the pix2pix model outperformed the U-Net (which produced unsatisfactory results in off-tissue regions) in terms of MI and SSIM, as shown in Table \ref{tab1}, with white padding achieving superior MI and very similar SSIM. Fig. \ref{fig1} shows two examples of synthetic (fake) histology images from the validation set, alongside their corresponding MSI and real histology images. The MSI images shown have a $m/z=885.55213$. Notably, Samples 1 (Fig. \ref{fig1}A-C) and 2 (Fig. \ref{fig1}D-F) show definite similarities in the overall tissue shape between real and fake. Furthermore, Sample 2's real histology (Fig. \ref{fig1}E) contains blue pen marks in the bottom-left and bottom-right corners, which have not been translated to the fake histology (Fig. \ref{fig1}F). Additionally, both samples show slightly darker colouration in correct parts.
\begin{table}[h]
\begin{center}
\caption{Comparison of experimental results for synthesising histological images using a U-Net and pix2pix models, where $B$ denotes black padding and $W$ white padding.}\label{tab1}
\setlength{\tabcolsep}{9pt} %Sets horizontal (column) spacing
\renewcommand{\arraystretch}{1.25} %Sets vertical (row) spacing
\begin{tabular}{|c||c|c|}
\hline
 & \textbf{MI} & \textbf{SSIM}\\
\hline
\hline
U-Net ($B$) &  0.429 & 0.530\\
\hline
pix2pix ($B$) &  1.349 & 0.949\\
\hline
pix2pix ($W$) & \textbf{1.353} & \textbf{0.949}\\
\hline
\end{tabular}
\end{center}
\end{table}
\section{Conclusion}
Our results show that image-to-image translation can synthesise histology images from MSI, which are qualitatively and quantitatively similar to real examples, serving as a promising solution for registration of the two modalities. The proposed pipeline utilises a pix2pix model which synthesises histological images using peak-picked MSI data, resulting in increases in MI and SSIM of $+0.924$ and $+0.419$, respectively, compared to a U-Net model. Visually, the results resemble their real counterpart's tissue shape and have minimal artifacts. Future work will focus on increasing the training dataset size and diversity, and comparing against state-of-the-art image-to-image translation models (e.g. Palette \cite{saharia2022}).
\section{Acknowledgments}
This work was supported by the Engineering and Physical Sciences Research Council [grant number EP/T518177/1].
\bibliographystyle{splncs04}
\end{document}